\newif\ifLinksBlack %black colored links
\LinksBlackfalse

\documentclass[conference]{IEEEtran}
\IEEEoverridecommandlockouts
\usepackage{cite}
\usepackage{amsmath,amsthm,amssymb,amsfonts}
\usepackage{algorithmic}
\usepackage{graphicx,subcaption}
\usepackage{textcomp}
\usepackage{xcolor}
\usepackage{epstopdf}
\usepackage{setspace}

\usepackage{psfrag}

\usepackage[linesnumbered,lined,boxed,commentsnumbered, ruled]{algorithm2e} %required for algorithms pseudocode

\def\BibTeX{{\rm B\kern-.05em{\sc i\kern-.025em b}\kern-.08em
		T\kern-.1667em\lower.7ex\hbox{E}\kern-.125emX}}

\ifLinksBlack
\RequirePackage[colorlinks,allcolors=black]{hyperref}
\else
\RequirePackage[colorlinks,allcolors=blue]{hyperref}
\fi

\graphicspath {{figures/}}

\newcommand{\I}{\mathbf{I}}
\newcommand{\IK}{\mathbf{I}_{K}}

\newcommand{\Q}{\mathbf{Q}}
\newcommand{\U}{\mathbf{U}}
\newcommand{\Wbf}{\mathbf{W}}
\newcommand{\Wbfh}{\mathbf{W}^{H}}

\newcommand{\Hbf}{\mathbf{H}}
\newcommand{\Hbfh}{\mathbf{H}^{H}}
\newcommand{\V}{\mathbf{V}}
\newcommand{\Z}{\mathbf{Z}}

\newcommand{\h}{\mathbf{h}}

\newcommand{\y}{\mathbf{y}}
\newcommand{\x}{\mathbf{x}}
\newcommand{\z}{\mathbf{z}}

\newcommand{\hatx}{\hat{\mathbf{x}}}
\newcommand{\hatn}{\hat{\mathbf{n}}}

\newcommand{\hatu}{\hat{\mathbf{u}}}
\newcommand{\Sbf}{\mathbf{\Sigma}}

\newcommand{\Mp}{M_{\text{p}}}
\newcommand{\Np}{N_{\text{p}}}

\DeclareMathOperator*{\argmax}{arg\,max}

\begin{document}
	
	\title{An Iterative Interference Cancellation Algorithm for Large Intelligent Surfaces}
	
	\author{
		\IEEEauthorblockN{
			Jes\'{u}s Rodr\'{i}guez S\'{a}nchez~\href{https://orcid.org/0000-0002-5531-1071}{\includegraphics[scale=0.04]{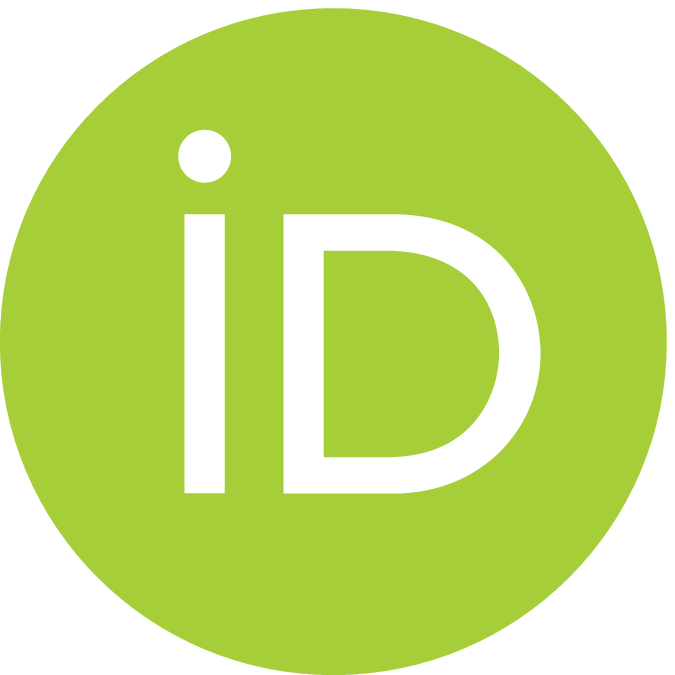}},
			Fredrik Rusek~\href{https://orcid.org/0000-0002-2077-3858}{\includegraphics[scale=0.04]{orcid.eps}},
			Ove Edfors~\href{https://orcid.org/0000-0001-5966-8468}{\includegraphics[scale=0.04]{orcid.eps}},
			and Liang Liu~\href{https://orcid.org/0000-0001-9491-8821}{\includegraphics[scale=0.04]{orcid.eps}}}\\
		\IEEEauthorblockA{Department of Electrical and Information Technology, Lund University, Sweden}
		\IEEEauthorblockA{\{jesus.rodriguez, fredrik.rusek, ove.edfors, and liang.liu\}@eit.lth.se}
	}
	
	\maketitle
	
	\begin{abstract}
		The Large Intelligent Surface (LIS) concept is a promising technology aiming to revolutionize wireless communication by exploiting spatial multiplexing at its fullest. Despite of its potential, due to the size of the LIS and the large number of antenna elements involved there is a need of decentralized architectures together with distributed algorithms which can reduce the inter-connection data-rate and computational requirement in the Central Processing Unit (CPU). In this article we address the uplink detection problem in the LIS system and propose a decentralize architecture based on panels, which perform local linear processing. We also provide the sum-rate capacity for such architecture and derive an algorithm to obtain the equalizer, which aims to maximize the sum-rate capacity. A performance analysis is also presented, including a comparison to a naive approach based on a reduced form of the matched filter (MF) method. The results shows the superiority of the proposed algorithm.
	\end{abstract}
	
	\section{Introduction}
\begin{figure}[t]
		\centering
		\includegraphics[width=\linewidth]{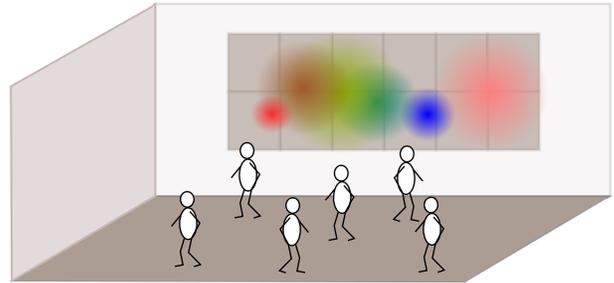}
		\caption{A LIS serving multiple users simultaneously.}
		\label{fig:LIS_concept}
		\vspace*{-4mm}
	\end{figure}

	We envision a future where man-made surfaces are electromagnetically
	active enabling wireless communication, wireless charging and remote sensing \cite{husha_data,husha_vtc,husha_pimrc}. The LIS concept is illustrated in Fig. \ref{fig:LIS_concept} where the LIS serves multiple users simultaneously. As it can be seen from the figure, the LIS is placed relatively close to the users. Thus parts of the LIS receive signals from different users with potential overlap if they are close to each other or in the same direction towards the surface. Users who are close to the LIS have a more concentrated and smaller $\textit{fingerprint}$ than the users who are far away, which is more spread as depicted in the figure. All parts of the surface are active, so they are able to transmit and receive electromagnetic waves with a certain control. The large aperture of the LIS allows to beamform in 3D space with high resolution, which helps to discriminate users even if their fingerprints overlap on the surface and therefore mitigating any potential inter-user interference. As pointed out in \cite{husha_data}, there is no practical difference between a continuous LIS and a grid of antennas (discrete LIS) as the surface area grows, provided that the antenna spacing is sufficiently dense. Based on this, we consider a discrete version of a LIS for a practical reason through the rest of this article.
	
	Even tough the potential of LIS is clear from theoretical point of view, there is a lack of algorithms specially designed for it. The large dimension and volume of antennas taking place makes traditional centralized methods not suitable for implementation due to excessive computational complexity and inter-connection data-rate required in the CPU. There is a need for distributed algorithms together with decentralized architectures.
	
	A preliminary work was published \cite{juan_VTC19} proposing a fully distributed algorithm for approximate zero-forcing (ZF) equalizer in LIS. Despite of achieving remarkable results in terms of signal-to-interference ratio (SIR), it requires multiple iterations of all processing nodes in the system with the subsequent simultaneous exchange of data between adjacent nodes, making difficult to pipeline it, in order to support simultaneous calculation of different equalizers for different physical-resource-blocks (PRB).
	
	In this paper we propose a decentralized architecture for LIS based on panels, which are connected to a backplane for data aggregation. The CPU receives the result of the backplane processing. Apart from this connection, panels are connected to each other by dedicated links forming a daisy-chain, which facilitates the pipelining of different equalizers. We also derive sum-rate capacity expression for this architecture, and based on that, we develop an algorithm for uplink detection which aims to maximize such capacity.
	
	We include performance analysis based on system parameters such as physical panel size and number of panels connections. We compared the proposed algorithm with a naive approach, consisting of a reduced version of the Matched Filter (MF) equalizer as reference baseline for our comparison.

	\section{System Model}
	\label{section:sys_model}

	\begin{figure}[ht]
		\footnotesize
		\centering
		\psfrag{xh}{$\hatx$}
		\psfrag{N}{$N$}
		\psfrag{K}{$K$}
		\psfrag{P}{$P$}
		\psfrag{MP}{$M_{p}$}
		\psfrag{NP}{$N_{p}$}
		\psfrag{z}{$\z$}
		\psfrag{y}{$\y$}
		\psfrag{x}{$\x$}
		\psfrag{W1}{$\Wbf_{1}$}
		\psfrag{W2}{$\Wbf_{2}$}
		\psfrag{WP}{$\Wbf_{P}$}
		\psfrag{H1}{$\Hbf_{1}$}
		\psfrag{H2}{$\Hbf_{2}$}
		\psfrag{HP}{$\Hbf_{P}$}
		\psfrag{Z1}{$\Z_{1}$}
		\psfrag{Z2}{$\Z_{2}$}
		\psfrag{ZP}{$\Z_{P-1}$}
		\psfrag{BACKPLANE}{$\text{BACKPLANE}$}
		\includegraphics[width=0.95\linewidth]{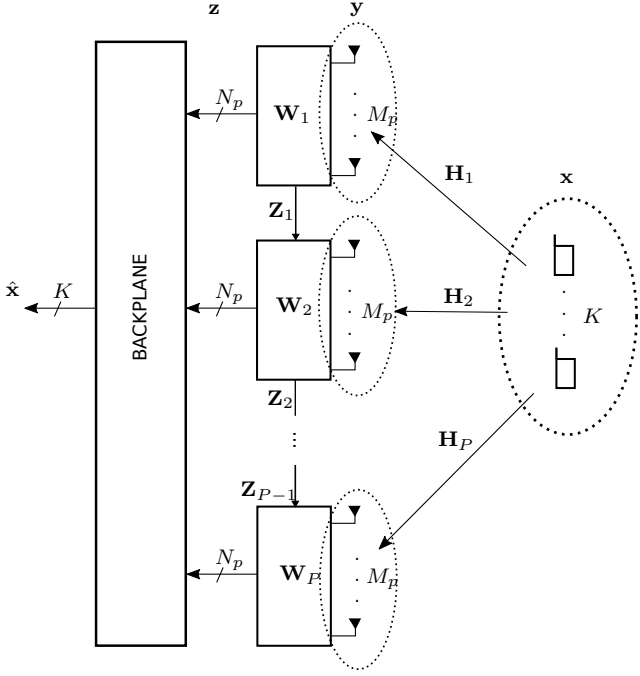}
		\caption{$K$ users transmitting to an M-elements discrete-LIS formed by $P$ panels.}
		\label{fig:scenario}
	\end{figure}
		
	We consider the transmission from $K$ single antenna users to a discrete LIS containing $M$ active antenna elements, as illustrated in Fig \ref{fig:scenario}. The LIS is divided into $P$ squared panels or LIS-units, each with $M_{\text{p}}$ elements, such that $M_{\text{p}} \cdot P = M$. Each panel has $N_p$ outputs. The total number of them is $N$, such that $N = N_{\text{p}} \cdot P$. These outputs are connected to a backplane, which collects and process the incoming data, and provides the CPU with $K$ values. In this article we consider the case $M \geq N \geq K$ (which implies $\Mp \geq \Np$), so the inter-connection data-rate is reduced from the antenna elements interface (vector $\y$ in the figure) to the backplane input ($\z$), and similarly to the CPU interface ($\hatx$). In this article we do not cover the design of the backplane, which is left for further work.
	
	The $M\times 1$ received vector at the LIS is given by
	\begin{equation}
	\mathbf{y} = \sqrt{\rho}\Hbf\mathbf{x}+\mathbf{n},
	\end{equation}
	where $\mathbf{x}$ is the $K\times 1$ user data vector, $\mathbf{H}$ is the $M \times K$ normalized channel matrix, such that $\|\Hbf\|^{2}=MK$, $\rho$ the $\mathrm{SNR}$ and $\mathbf{n} \sim \mathcal{CN}(0,\I)$ is a $M \times 1$ noise vector.
	
	Assuming the location of user $k$ is $(x_{k},y_{k},z_{k})$, where the LIS is in $z=0$. The channel between this user and a LIS antenna at location $(x,y,0)$ is given by the complex value \cite{husha_data}
	\begin{equation}
	h_{k}(x,y)=\frac{\sqrt{z_{k}}}{2\sqrt{\pi} d_{k}^{3/2}}\exp{\left( -\frac{2\pi j d_{k}}{\lambda} \right)},
	\label{eq:channel}
	\end{equation}
	where $d_{k}=\sqrt{z_{k}^{2}+(x_{k}-x)^2+(y_{k}-y)}$ is the distance between the user and the antenna, and Line of Sight (LOS) propagation between them is assumed. $\lambda$ is the wavelength.
	
	The channel matrix can be expressed as
	\begin{equation}
	\Hbf=[\Hbf_{1}^{T},\Hbf_{2}^{T},\cdots \Hbf_{P}^{T}]^{T},
	\label{eq:H_structure}
	\end{equation}
	where $\Hbf_{i}$ is the $\Mp \times K$ channel matrix of the $i$-th panel.
	We assume each panel has perfect knowledge of its local channel.
		
	\section{Uplink detection}
	\label{section:uplink_detection}
	
	LIS performs a linear filtering on the incoming signal in the panels such as
	\begin{equation}
	\mathbf{z}= \Wbf^{H} \y = \sqrt{\rho}\Wbfh\Hbf \mathbf{x} + \hatn,
	\end{equation}
	where $\Wbf^{H}$ is the $N \times M$ equalization-filter matrix and $\hatn=\Wbfh\mathbf{n}$ is the filtered noise.	
	
	\subsection{Sum-Rate Capacity}
	The mutual information between $\z$ and $\x$ is $I(\x;\z)=H(\z)-H(\z|\x)$. Assuming gaussian signaling transmitted by users, the mutual information for a given $\Hbf$ and $\Wbf$ can be further expanded as
	\begin{equation}
	\begin{split}
	I(\x;\z) &= \log_{2}|\Sbf_{\z\z}| - \log_{2}|\Sbf_{\hatn\hatn}|\\
	&= \log_{2} |\rho\Wbfh\Hbf\Hbfh\Wbf+\Wbfh\Wbf|\\
	&-\log_{2} |\Wbf^{H}\Wbf|,
	\end{split}
	\label{eq:I(x;z)}
	\end{equation}
	where $\Sbf_{\z\z}$ and $\Sbf_{\hatn\hatn}$ are the covariance of the multivariate complex gaussian vector $\z$ and $\hatn$ respectively.
	If $\Wbf$ is full-rank matrix, and taking into account that $M \geq N$, then $(\Wbfh \Wbf)^{-1}$ exists and we can rewrite \eqref{eq:I(x;z)} as
	\begin{equation}
	\begin{split}
	I(\x;\z) &= \log_{2} |\IK + \rho\Hbfh\Wbf(\Wbfh\Wbf)^{-1}\Wbfh\Hbf|,\\
	\end{split}
	\label{eq:I(x;z)2}
	\end{equation}
	thus the sum-rate capacity at $\z$ interface is given by
	\begin{equation}
	C_{\z} = \max_{\Wbf} \log_{2} | \IK + \rho \Hbfh \Wbf (\Wbfh \Wbf)^{-1} \Wbfh \Hbf |.\\
	\label{eq:Cz}
	\end{equation}

	\subsection{Single panel case}
	First we analyze the simple case where there is one panel in the system ($P=1$), and then $N=\Np$ and $M=\Mp$.
	
	Let $\U_{W} \Sbf_{W} \V_{W}^{H}$ and $\U_{H} \Sbf_{H} \V_{H}^{H}$ be the singular value decomposition (SVD) of $\Wbf$ and $\Hbf$ respectively, then \eqref{eq:Cz} can be formulated as
	\begin{equation}
	C_{\z} = \max_{\U_{W}} \log_{2} | \IK + \rho \Hbfh \U_{W} \hat{\I} \U_{W}^{H} \Hbf |,\\
	\label{eq:C2}
	\end{equation}
	where $\hat{\I}$ is a $\Mp \times \Mp$ diagonal matrix defined as
	\begin{equation}
	\hat{\I}_{i,i} = \begin{cases}
	1 & \text{if } i \leq \Np\\
	0 & \text{if } \Np < i \leq \Mp.\\
	\end{cases}
	\label{eq:hatI}
	\end{equation}

	From \eqref{eq:C2} it can be seen that only the $\Np$-first left singular vectors of $\Wbf$ are relevant from capacity point of view. Infinite number of matrices $\Wbf$ fulfill this criteria and therefore provide the same capacity. Among those, we can select $\Wbf_{\text{SP}}=\widetilde{\U}_{H}$ for simplicity, where $\widetilde{\U}_{H}$ is a $\Mp \times \Np$ semi-unitary matrix made by the $N$-first left singular vectors of $\Hbf$.
	
	\subsection{Multiple panels case}
	In the multiple-panels case ($P>1$) we consider the panelized architecture shown in Fig. \ref{fig:scenario}, where each panel performs local processing on the received signal and delivers the result to the backplane. In this case the filter matrix $\Wbf$ has the following structure
	\begin{equation}
	\Wbf = \begin{bmatrix}
	\Wbf_{1}   & \mathbf{0} & \cdots  & \mathbf{0}\\
	\mathbf{0} & \Wbf_{2}   & \cdots  & \mathbf{0}\\
	\mathbf{0} & \mathbf{0} & \cdots  & \mathbf{0}\\
	\vdots	   & \vdots		& \ddots  & \vdots\\
	\mathbf{0} & \mathbf{0} & \cdots  & \Wbf_{P}
	\end{bmatrix},
	\label{eq:W_structure}
	\end{equation}
	where $\Wbf_{i}$ is the $\Mp \times \Np$ matrix filter of the $i$-th panel.
	
	The sum-rate capacity for this architecture is given by the general expression \eqref{eq:Cz}. However, if we take into account the block structure of $\Hbf$ and $\Wbf$ presented in \eqref{eq:H_structure} and \eqref{eq:W_structure} respectively, \eqref{eq:Cz} can be written in a more specific form as
	\begin{equation}
	\begin{split}
	C_{\z} &= \max_{\{\Wbf_{i}\}} \log_{2} | \IK + \rho \sum_{i=1}^{P} \Hbf_{i}^{H} \Wbf_{i} (\Wbf_{i}^{H} \Wbf_{i})^{-1} \Wbf_{i}^{H} \Hbf_{i} |\\
	&= \max_{\{\mathbf{S}_{i}\}} \log_{2} | \IK + \rho \sum_{i=1}^{P} \Hbf_{i}^{H} \mathbf{S}_{i} \Hbf_{i} |,\\
	\end{split}
	\label{eq:capacity_multi}
	\end{equation}
	where $\mathbf{S}_{i} \triangleq \Q_{i} \Q^{H}_{i}$ and $\Q_{i}$ is a $\Mp \times \Np$ semi-unitary matrix, and consequently follows that $\Wbf_{i}=\Q_{i}$ similarly to the single-panel case. For the last expression in \eqref{eq:capacity_multi} it is assumed that all $\Wbf_{i}$ are full-rank so the inverse exists.
	
	\section{ALGORITHMS}
	
	In this section we proposed two algorithms to obtain the $P$ filtering matrices $\{\Wbf_{i}\}$. The first one is a naive approach with relatively low computational complexity based on the known matched filter method, which we select conveniently as a comparison baseline for our proposed algorithm, which is presented later on in this section.
	
	\subsection{Reduced Matched Filter (RMF)}
	RMF consists of a reduced version of the known Matched Filter (MF) method. In this case, only the $N_{\text{p}}$ $\textit{strongest}$ columns of $\Hbf_{i}$ are used to formulate $\Wbf_{i}$. The $\textit{strenght}$ of a column $\h_{n}$ is defined as $\|\h_{n}\|^{2}$. The $\Mp \times \Np$ filtering matrix of the $i$-th panel is then expressed as
	\begin{equation}
	\Wbf_{\text{RMF},i} = \left[ \h_{k_1}, \h_{k_2}, ...,  \h_{k_{Np}} \right],
	\label{eq:W_RMF}
	\end{equation}
	where $\h_{n}$ is the $\Mp \times 1$ channel vector for the $n$-th user, $\{k_{i}\}$ represents the set of indexes relative to the $N_{\text{p}}$ strongest users.
	
	\subsection{Iterative Interference Cancellation (IIC) Algorithm}
	The IIC algorithm aims to solve the optimization problem described in \eqref{eq:capacity_multi}. It is an iterative algorithm based on a variant of the known multiuser water-filling method \cite{Yu}. The algorithm is described in Algorithm \ref{algo:MUWF} using pseudocode. The main differences with regard to the optimization problem described in \cite{Yu} is that in our case there is no power constraint (because $\Wbf_{i}$ are intended to be used in uplink) and the matrices $\mathbf{S}_{i}$ consist of an outer product of two semi-unitary matrices. 

	\IncMargin{1em}
	\begin{algorithm}[ht]
		\SetKwInOut{Input}{Input}
		\SetKwInOut{Output}{Output}
		\SetKwInOut{Preprocessing}{Preprocessing}
		\SetKwInOut{Init}{Init}
		\Input{$\{\Hbf_{i}\}, i=1 \cdots P$}
		\Preprocessing{ $\mathbf{S}_{i} = \mathbf{0}_{Mp}, \forall i$}
		\For{$i = 1,2,...,P$}{
			$\mathbf{Z}_{i-1} = \IK + \rho \sum_{j=1,j\neq i}^{P} \Hbf_{i}^{H} \mathbf{S}_{i} \Hbf_{i}$\\
			
			$\mathbf{S}_{i} = \argmax_{\mathbf{S}} |\rho \Hbf_{i}^{H} \mathbf{S} \Hbf_{i} + \mathbf{Z}_{i-1}|$

		}
		\caption{IIC algorithm pseudocode}
		\label{algo:MUWF}
		\Output{$\{\mathbf{S}_{i}\}, i=1 \cdots P$}
	\end{algorithm}\DecMargin{1em}
	At each iteration of the algorithm, one panel filtering matrix is obtained, so $P$ iterations are required. In principle, once the last iteration is complete, the algorithm can start again updating the resulting $\{\mathbf{S}_{i}\}$ until a convergence criteria is achieved, but we consider one iteration in this work for simplicity.
	
%	[connect optimization problem with eq.8]
	
	\section{PROCESSING DISTRIBUTION}
	
	For each iteration of Algorithm \ref{algo:MUWF}, knowledge of all $\{\Hbf_{i}\}$ and $\{\mathbf{S}_{i}\}$ is required.	
	One way of achieving this is by executing the algorithm in the CPU. In this case, the CPU requires CSI from all panels, which consist of $M \times K$ complex numbers, increasing the inter-connection data-rate in the backplane and computational demand from the CPU point of view.
	
	Another way consists of executing the algorithm IIC in a decentralized form, by running each iteration in a different panel. In this case, the matrix $\mathbf{Z}_{i-1}$ is a message received by the previous adjacent panel, and contains information from all previous panels. Matrix $\Hbf_{i}$ is known only locally at $i$-th panel. The matrix $\Z$ is passed from panel to panel using the dedicated connections depicted in Fig. \ref{fig:scenario}.

	\IncMargin{1em}
	\begin{algorithm}[ht]
		\SetKwInOut{Input}{Input}
		\SetKwInOut{Output}{Output}
		\SetKwInOut{Preprocessing}{Preprocessing}
		\SetKwInOut{Init}{Init}
		\Input{$\{\Hbf_{i}, \Z_{i-1}\}$}
		$\mathbf{S}_{i} = \argmax_{\mathbf{S}} |\rho \Hbf_{i}^{H} \mathbf{S} \Hbf_{i} + \mathbf{Z}_{i-1}|$\\
	
		$\mathbf{Z}_{i+1} = \mathbf{Z}_{i} + \Hbf_{i}^{H} \mathbf{S}_{i} \Hbf_{i}$
		\caption{IIC algorithm in $i$-th panel}
		\label{algo:IIC_i}
		\Output{$\{\mathbf{S}_{i}, \mathbf{Z}_{i}\}$}
		
	\end{algorithm}\DecMargin{1em}

	The pseudocode of the algorithm for the $i$-th panel is illustrated in Algorithm \ref{algo:IIC_i}. As it can be observed, IIC divides the optimization problem defined in \eqref{eq:capacity_multi} in $P$ smaller and local optimization problems as follows
	\begin{equation}
	\mathbf{S}_{i} = \argmax_{\mathbf{S}} \log_{2} | \rho \Hbf^{H}_{i} \mathbf{S} \Hbf_{i} + \mathbf{Z}_{i-1}|,\\
	\label{eq:Si_multi_argmax}
	\end{equation}
	The solution to \eqref{eq:Si_multi_argmax} is $\mathbf{S}_{i}=\hat{\Q}_{i}\hat{\Q}_{i}^{H}$, where $\hat{\Q}_{i}=[\hatu_{1},\hatu_{2}, \cdots, \hatu_{\Np}]$ and $\hatu_{n}$ is the $n$-th left-singular vector of $\hat{\Hbf}_{i}=\sqrt{\rho}\Hbf_{i} \U_{z} \Sbf_{z}^{-1/2}$ corresponding to the n-$th$ ordered singular value, and $\mathbf{Z}_{i-1}$ defined as in Algorithm \ref{algo:IIC_i}, whose eigen-decomposition is $\U_{z} \Sbf_{z} \U_{z}^{H}$. It can be shown that algorithm IIC increases monotonically the sum-rate capacity at each iteration (see Appendix).
		
	\section{Results}
	In this section we analyze the performance of the proposed algorithm in a $1m \times 10m$ LIS located in a $3m$ (height) $\times 30m$ (width) $\times 30m$ (depth) area. The LIS is placed in the center of one of the sides. We consider two possible panel sizes, one small ($20cm \times 20cm$) and one large ($1m \times 1m$), an study the impact of this parameter in the performance. Then, the number of panels, $P$, is $250$ and $10$ respectively, and the number of antennas per panel, $\Mp$, is $16$ and $400$ respectively. 20 users are uniformly located in the mentioned area. Sufficient channel realizations are generated (each with random user location) for each experiment. The wavelength $\lambda$ is 5cm and the antenna spacing is $\lambda/2$ in all cases.
		
\begin{figure}%[ht]
		\footnotesize
		\centering
		\includegraphics[width=0.73\linewidth]{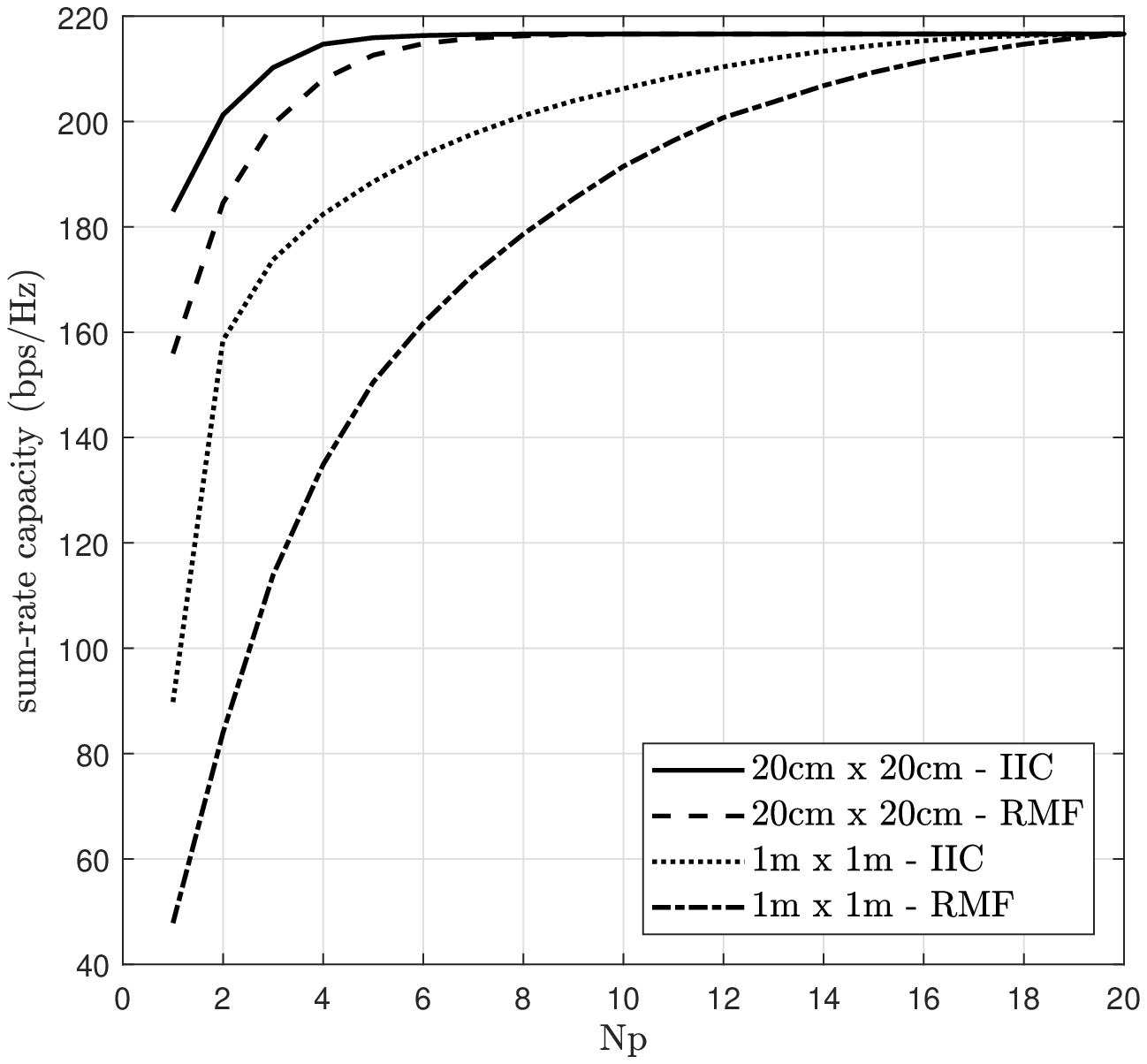}
		\caption{Average sum-rate vs. $\Np$.}
		\label{fig:sum_rate_vs_Np}
		\vspace*{-4mm}
	\end{figure}

	Average sum-rate capacity at interface $\z$, $C_{\z}$, versus $\Np$ is depicted in Fig. \ref{fig:sum_rate_vs_Np} for both panel sizes and both algorithms. It can be observed that the sum-rate grows as $\Np$ increases in all cases, converging to the same value, which is the average channel capacity for this configuration. It can also be observed as the LIS with small-panels converges earlier to channel capacity, meaning that each panel needs less number of outputs in the small-panels case for same target performance. It is also clear from the figure that the IIC algorithm provides better performance than RMF for both panel sizes and all tested $\Np$ values. This can be used in two forms: for a certain $\Np$ IIC achieves a performance gain compared to RMF, and for a certain target performance results in a reduction of $\Np$ compared to RMF case.

\begin{figure}[ht]
		\footnotesize
		\centering
		\includegraphics[width=0.73\linewidth]{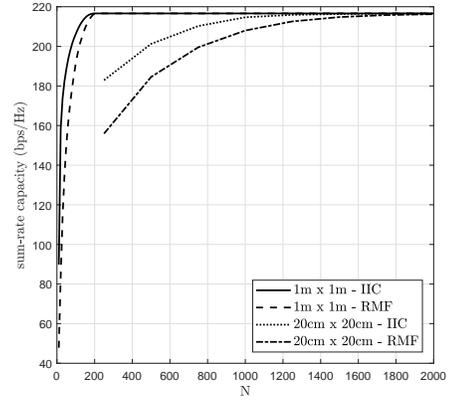}
		\caption{Average sum-rate vs. $N$.}
		\label{fig:sum_rate_vs_N}
		\vspace*{-4mm}
	\end{figure}

	However, even tough the required number of outputs is lower in the small-panels case (for a target performance), the larger number of panels needed in that case makes the total number of outputs compromised when compared to the large-panels case. In Fig. \ref{fig:sum_rate_vs_N} we analyze this point by showing the average sum-rate $C_{\z}$ versus the total number of outputs (N) for both panel sizes and both algorithms. It can be seen as for the same N value, large-panels achieve better performance, at the cost of higher number of outputs per panel.
		
	\section{Conclusions}
	In this article we present a large Intelligent Surface (LIS) architecture and propose one algorithm for uplink data detection. For comparison purposes we also introduce a naive method based on the known Matched Filter. The architecture is based on panels connected to a backplane for data aggregation, and a CPU which receives and process the aggregated data. The presented algorithms are executed locally in the panels. The result of such processing is delivered to the backplane using a certain number of outputs. We analyze and compare the performance of each algorithm versus the total number such outputs, and the ones per panel, identifying interesting trade-offs and design guidelines. From the comparison it is clear that the proposed algorithm achieves better performance than the naive approach at the cost of dedicated panel-to-panel connections.
	
	\section{Appendix}
	\label{section:appendix}
	
	Let us define the sum-rate capacity achieved by Algorithm \ref{algo:MUWF} up to $i$-th panel as $C_{\z,i}$, then
	\begin{equation}
	\begin{split}
	C_{\z,i} &= \log_{2} |\rho \Hbf_{i}^{H} \mathbf{S}_{i,opt} \Hbf_{i} + \mathbf{Z}_{i-1}| \\
	&= \log_{2}|\rho \Hbf_{i}^{H} \mathbf{S}_{i,opt} \Hbf_{i} + \U_{z} \Sbf_{z} \U_{z}^{H}|\\
	&= \log_{2}|\Sbf_{z}| |\rho \Sbf_{z}^{-1/2} \U_{z}^{H} \Hbf_{i}^{H} \mathbf{S}_{i,opt} \Hbf_{i} \U_{z} \Sbf_{z}^{-1/2} + \I| \\
	&= C_{\z,i-1} + \log_{2}|\hat{\Hbf}_{i}^{H} \mathbf{S}_{i,opt} \hat{\Hbf}_{i} + \I|,
	\end{split}
	\nonumber
	\end{equation}
	where $\hat{\Hbf}_{i} = \sqrt{\rho} \Hbf_{i} \U_{z} \Sbf_{z}^{-1/2}$. Using the proposed solution for $\mathbf{S}_{i}$ the second term si always positive, so the sum-rate capacity increases at each iteration.
	
	\bibliographystyle{IEEEtran}
	\bibliography{IEEEabrv,references}
	\nocite{book:IT}
	
\end{document}